# Bridging Optical Sensing and Wearable Health Monitoring: A Functionalized Plasmonic Nanopillar for Non-Invasive Sweat Glucose Detection

Ling Liu[1], Kuo Zhan[1]*, Joni Kilpijärvi[2], Matti Kinnunen[2], Yuan Zhang[1], Mulusew Yaltaye[1], Yang Li[1], Artem Zhyvolozhnyi[3], Anatoliy Samoylenko[3], Seppo Vainio[3], Jianan Huang[1]*

[1]Research Unit of Disease Networks, Faculty of Biochemistry and Molecular Medicine University of Oulu, Oulu, Finland.

[2]Polar Electro Oy, Oulu, Finland.

[3]Faculty of Biochemistry and Molecular Medicine, Disease Networks Research Unit, InfoTech Oulu, Kvantum Institute, University of Oulu, Borealis Biobank of Northern Finland, Univ. Oulu Hospital, FI-90014 Oulu, Finland.

**Abstract.**

Continuous glucose monitoring (CGM) is vital for diabetes care, but current systems rely on invasive implants or electrochemical sensors that often cause discomfort and skin irritation. Non-invasive alternatives remain limited by low sensitivity and poor compatibility with complex sweat environments, highlighting the urgent need for a comfortable and reliable solution. Here, we report the development of a wearable optical sensor watch that integrates surface plasmon resonance (SPR) technology with a functionalized silver-coated silicon nanowire (Ag/SiNW) substrate for real-time, non-invasive glucose monitoring in sweat. The nanostructured sensor is functionalized with 4-mercaptophenylboronic acid (4-MPBA), enabling selective glucose capture and optical signal transduction through both Raman scattering and SPR shift. The dual-mode detection strategy was systematically optimized, and a miniaturized SPR system operating at 638 nm was successfully integrated into a wearable watch format with wireless data transmission to a mobile application. This wearable device demonstrated excellent sensitivity (LOD down to 0.12 mM) and high selectivity in detecting glucose within physiological sweat concentration ranges. Human subject trials confirmed its applicability in real-life scenarios. This study offers a promising non-invasive alternative to traditional CGM and highlights the potential of integrating nanophotonic sensors with wearable platforms for continuous health monitoring and personalized medicine.

# 1. Introduction

Diabetes mellitus affects over 537 million individuals worldwide and remains one of the most prevalent chronic diseases and leading causes of mortality[1-3]. Effective management of the disease requires frequent monitoring of blood glucose levels to avoid complications and maintain metabolic balance[4, 5]. Although continuous glucose monitoring (CGM) technologies have significantly advanced diabetes care, most existing systems rely on subcutaneous electrochemical sensors[6]. These minimally invasive approaches often result in discomfort, increased risk of infection, and reduced patient compliance, particularly during long-term use. Consequently, there is an increasing need for a non-invasive, reliable, and user-friendly glucose monitoring strategy suitable for prolonged daily applications.

Sweat, an easily accessible and information-rich body fluid, has emerged as a promising candidate for non-invasive biomarker monitoring[7, 8]. It contains various physiologically relevant analytes, including glucose, electrolytes, and lactate, and its composition correlates with blood chemistry under specific conditions[9, 10]. The presence of abundant sweat glands throughout the human skin enables straightforward and continuous sample collection[11, 12]. However, the relatively low and fluctuating concentration of glucose in sweat presents significant challenges for achieving adequate sensitivity and specificity. Additional issues such as interference from other sweat components, inconsistent secretion rates, and the complex conditions on the skin surface further complicate reliable glucose detection.

Several sensing strategies have been investigated to address these obstacles. Electrochemical sensors are commonly employed due to their compact form and rapid response[13]; however, they may cause irritation to the skin and exhibit poor selectivity in complex biological matrices. Optical methods such as colorimetry[14, 15] and fluorescence[16] offer improved sensitivity but often require bulky components and external light sources, which limit their suitability for integration into wearable formats. Surface plasmon resonance (SPR), among other optical techniques, provides real-time, label-free detection with high specificity[17, 18]. Nevertheless, SPR-based approaches remain largely restricted to laboratory settings, primarily due to challenges in miniaturization, fluidic stability, and integration with wearable systems.

In this work, a wearable SPR-based glucose sensor was developed that incorporates silver-coated silicon nanowires (Ag/SiNWs) functionalized with 4-mercaptophenylboronic acid (4-MPBA), as shown in **Schematic 1A**. The 4-MPBA molecules act as selective glucose receptors, enabling transduction of binding events into optical signals through changes in the local refractive index. The nanowire array structure enhances plasmonic coupling and signal amplification. This platform supports both Raman scattering and SPR-based detection, offering a dual-mode analytical capability. A compact laser-based optical system operating at 638 nm was integrated into a wearable watch, and wireless data transmission to a mobile application was implemented for real-time monitoring, as shown in **Schematic 1B**. The developed device enables accurate quantification of glucose in sweat within physiologically relevant ranges, achieving a detection limit as low as

0.12 mM. In vivo validation using human volunteers confirmed its practicality and robustness for continuous, non-invasive glucose tracking. This work presents a significant advance toward the realization of wearable photonic devices for real-time health monitoring and offers new opportunities for personalized diabetes care.

## 2. Materials and Methods

2.1 Material

The silicon wafer was Boron-doped p-type with a resistivity of 5 – 20 $\Omega \cdot cm^{-2}$, D-(+)-Glucose and 4-Mercaptobenzenepropanol Acid (4-MPBA) were purchased from Sigma-Aldrich Co., LLC. Polystyrene nanospheres of 500 nm in diameter was brought from Scientific fisher.

2.2 Fabrication of 4-MPBA - Au/SiNWs and Ag/SiNWs.

The silicon nanowire (SiNWs) array was improved based on previous research[19]. Briefly, polystyrene nanospheres with a diameter of 500 nanometers and a mixed solution of ethanol and ethylene glycol in a volume ratio of 1:2:1 was slowly injected onto the water surface to form a monolayer. A clean silicon wafer, previously immersed, was slowly raised to the water surface so that the polystyrene monolayer was on top of it. After air drying, the as-made substrate was placed in a reactive ion machine (Plasmalab80Plus RIE) and the cylindrical nanospheres were etched for 3 minutes and 15 seconds at 30 m Torr pressure, 30 W oxygen power and 5 W argon power. A 2 nanometers titanium layer and a 5 nanometers gold layer were then deposited on the substrate using a Sputtering Coating Machine (Q150T ES) and placed in a mixed solution of ethanol and pure water in a volume ratio of 1:1 and ultrasonicated for five minutes. The SiNWs array was prepared by immersing the substrate in an etchant consisting of 20 ml of water, 1 ml of 48% HF and 0.2 ml of 35% $H_2O_2$ for 10 minutes.

For the gold-coated silicon nanowire arrays (Au/SiNW): A Sputtering Coating Machine (Q150T ES) was used to deposit a layer of gold with a thickness of about 50 nm on the prepared SiNWs array. Finally, the prepared Au/SiNWs array was placed in a 6 mM 4-MPBA solution at room temperature overnight.

For the silver-coated silicon nanowire arrays (Ag/SiNWs): A Sputtering Coating Machine (Q150T ES) was used to deposit a layer of silver with a thickness of about 50 nm on the prepared SiNWs array. Finally, the prepared Ag/SiNWs array was placed in a 6 mM 4-MPBA solution at room temperature overnight.

2.3 Sweat collection

Four healthy volunteers provided written informed consent. The sweat collection area was gently cleaned with soap and water, and the sweat was collected while the volunteers sweated during exercise.

2.4 SERS measurement

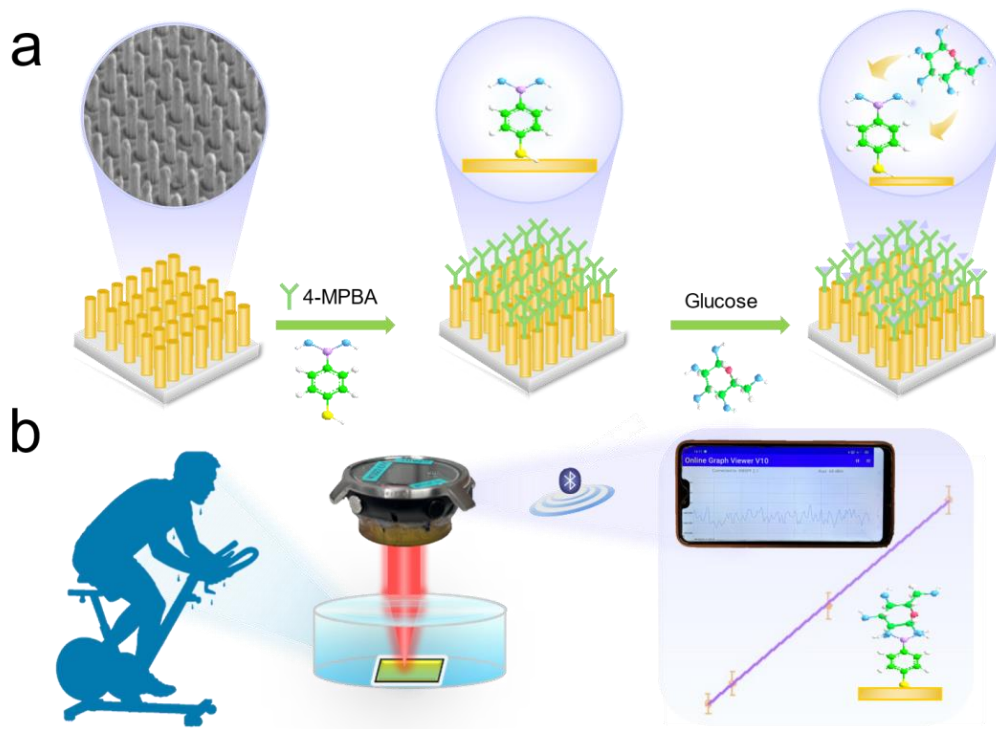

**Schema 1** (A) Flowchart of functionalizing nanopillars using 4-MPBA, (B) Schematic diagram of polar watch detecting glucose sensor.

Use Thermo Scientific Raman spectrometer to collect the Raman spectrum of glucose. Place the functionalized 1×1cm gold nanopillar in a glucose solution with a configured standard concentration. After waiting for 15 minutes, detect at different positions.

2.5 SPR measurement.

Raman spectra of pure glucose solutions and sweat samples were collected using a surface plasmon resonance (SPR) spectrometer, and functionalized 1×1 cm gold or silver nanopillars were placed in a concentration solution configured to a standard concentration and then diluted and assayed, and the assay was maintained for 15 minutes.

## 3. Results and discussion

3.1 Sensitivity and Specificity in Complex Sweat Environments

Accurate detection of glucose in sweat remains a fundamental challenge due to its low concentration and the presence of numerous interfering substances. Unlike blood, where glucose levels are relatively stable and abundant, sweat contains glucose in the micromolar to sub-millimolar range[20], often fluctuating with hydration and physical activity. Furthermore, the Raman scattering cross-section of glucose is intrinsically low[21-23], making signal acquisition extremely difficult, especially in the presence of biological background noise[24]. To address this issue, we employed 4-mercaptophenylboronic acid (4-MPBA) as a functionalization agent on gold- and silver-coated silicon nanowire (SiNW) arrays (**Fig.1a**). The boronic acid moiety binds selectively binds to cis-diol groups on glucose molecules, enhancing both the affinity and specificity of the sensor[24, 25]. When integrated with the nanowire structure, this configuration created an optically active interface capable of amplifying weak signals via both surface-enhanced Raman scattering (SERS) and surface plasmon resonance (SPR).

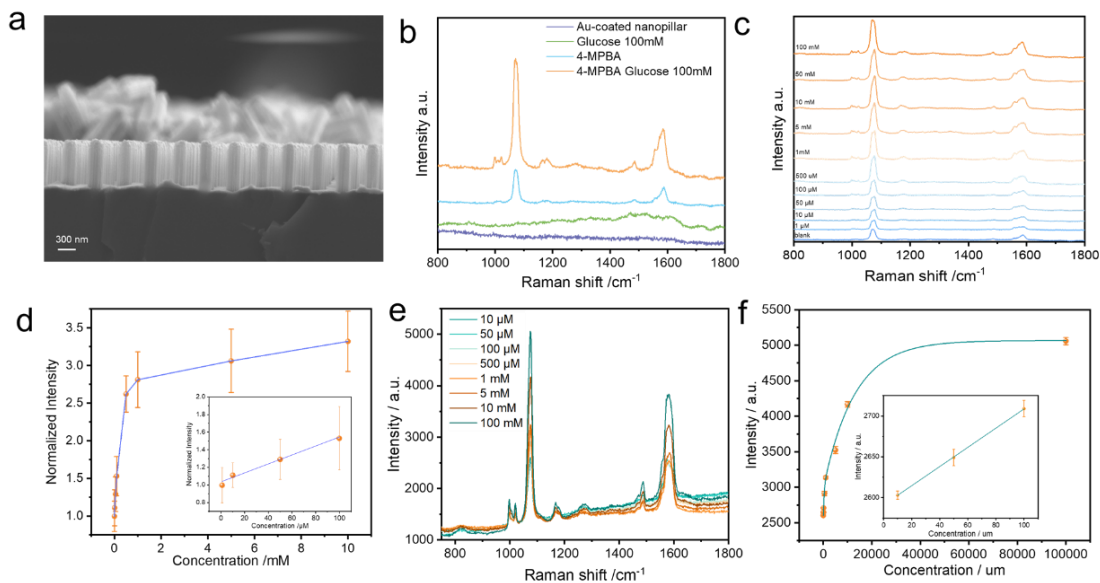

**Fig. 1.** (a) SEM of gold nanopillar from a side angle. (b) Comparison of Raman spectra of gold nanopillar (purple line), functionalized gold nanopillar (blue line), 100 mM glucose solution (green line), and functionalized gold nanopillar in glucose solution (orange line). (c) Raman spectra of 4-MPBA-functionalized Au-coated nanopillars versus glucose concentrations. (d) Calibration curve of normalized Raman intensity of 4-MPBA functionalized gold-coated nanorods versus glucose concentration at the 1073 cm$^{-1}$ peak. (e) Portable Raman spectra of 4-MPBA functionalized Au nanopillar versus glucose concentration. (f) Calibration curve of normalized Raman intensity of 4-MPBA functionalized gold-coated nanorods versus glucose concentration at the 1073 cm$^{-1}$ peak.

Our results demonstrate that the 4–MPBA–modified nanostructure generated a distinct Raman peak at 1073 cm$^{-1}$ corresponding to the C–C stretching of the aromatic ring, which exhibited a significant intensity increase in the presence of glucose[24]. As shown in **Fig. 1b**, the comparison between bare Au/SiNW, functionalized Au/SiNW, and glucose-exposed samples clearly confirms the molecular recognition capability and enhancement effect of the sensing interface. Signal intensity exhibited a clear dependence on glucose concentration in the low micromolar range, as illustrated in **Fig. 1c–f**, achieving a limit of detection (LOD) as low as 117.9 μM using Raman spectroscopy and 0.12 mM with the portable Raman. These findings indicate that the combination of molecular recognition via 4-MPBA and nanostructure-assisted optical enhancement effectively overcomes the long-standing barrier of selectivity and sensitivity in sweat-based glucose sensing. This molecular interface serves as a robust platform for reliable detection in biologically relevant concentration ranges, laying the foundation for practical non-invasive applications.

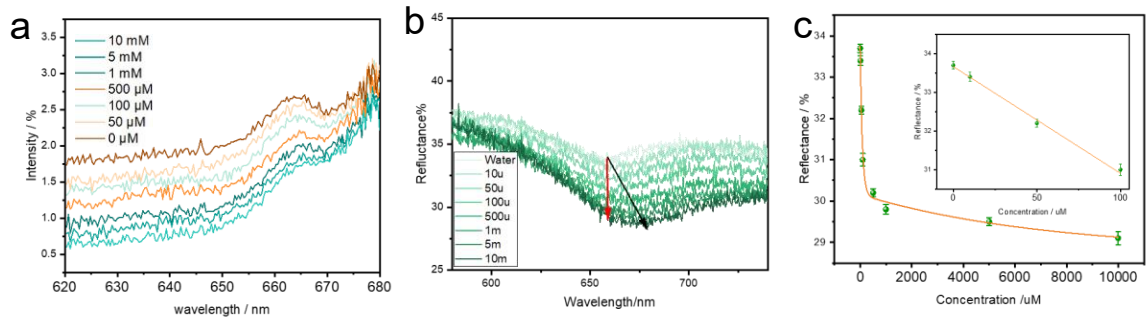

**Fig. 2** (a) SPR reflectance intensity of the functionalized Au-coated nanopillar versus the concentration of glucose. (b) SPR reflectance intensity of the functionalized Ag-coated nanopillar versus the concentration of glucose. (c) Linear relationship between SPR reflectance intensity of 4–MPBA–modified Ag-coated nanopillar and glucose concentration.

3.2 Material Optimization for Enhanced Red-Light SPR Performance

Surface plasmon resonance (SPR) techniques are highly sensitive to refractive index changes at metal–dielectric interfaces and have been widely applied in biomolecular detection. However, traditional SPR configurations often rely on gold thin films and prism-based optical setups, which exhibit suboptimal resonance in the red to near-infrared range. This spectral region is particularly desirable for wearable sensing applications due to the availability of compact, low-power red lasers diodes and deeper tissue penetration. Yet, gold films typically exhibit broad and weak SPR features at these wavelengths (**Fig. 2a, Fig. S3**), limiting their effectiveness in portable systems. To address this limitation, we explored material substitutes by replacing the conventional gold layer with a silver coating on the silicon nanowire substrate. Compared to gold, silver exhibits sharper and more intense plasmonic resonances in the visible and red regions due to its lower imaginary permittivity and higher plasmonic quality factor. Scanning electron microscopy (SEM) revealed that the silver-coated nanowires retained uniform morphology with a roughened surface structure, as shown in **Fig. S2**, which is favorable for both functionalization and field enhancement.

SPR measurements performed using red light (620–680 nm) confirmed a marked improvement in resonance sharpness and sensitivity when using Ag/SiNW substrates. As shown in **Fig. 2b**, the reflectance spectra showed distinct red shifts in response to increasing glucose concentration, with well-separated peaks and minimal baseline drift. The silver-based substrate enabled a clear linear response in the low micromolar glucose range, with a limit of detection improved to 0.02 mM (**Fig. 2c**). This material optimization is particularly significant for wearable system design, as it enables the use of red laser diodes that are compact, low-cost, and energy-efficient. By achieving strong SPR responses within this spectral window, our sensor architecture meets both the optical and practical requirements for integration into miniaturized, battery-powered platforms.

3.3 System Integration and Real-Time Performance of the Wearable Watch

While many optical sensing technologies offer high analytical performance in controlled laboratory environments, their translation into practical, real-time applications remains limited by issues such as device complexity, power consumption, and lack of user-friendly interfaces. To overcome these limitations, we developed a fully integrated wearable watch system (Polar Watch) that incorporates a compact 638 nm laser, a 4-MPBA modified silver-coated SiNW SPR sensor, and wireless communication modules for mobile device connectivity. The selection of the 638 nm wavelength was informed by our prior SPR optimization, ensuring maximal plasmonic response while maintaining low power consumption. The sensing module was miniaturized and embedded within the body of the watch, allowing for non-invasive monitoring from the skin surface. During operation, the reflected light intensity was continuously recorded and transmitted in real time to a custom smartphone application, enabling users to visualize glucose fluctuations without external instruments. In calibration experiments using standard glucose solutions, the system exhibited a clear and progressive decrease in reflectance with increasing glucose concentration, confirming consistent sensor response across a physiologically relevant range (**Fig. 3a–g**). More importantly, the relative intensity changes demonstrated a linear relationship in the 0–500 μM range, corresponding to the typical glucose levels found in human sweat. The calculated detection limit of 0.052 mM aligns well with reported concentrations in healthy individuals and diabetic patients, suggesting practical applicability for real-world use. By successfully embedding the optical sensor into a wearable watch with wireless data transmission, our system bridges the gap between high-performance biochemical sensing and user-oriented health technology. This level of integration demonstrates not only technical feasibility but also strong potential for daily glucose management in non-clinical settings.

3.4 Real-World Validation Using Human Sweat Samples

To evaluate the practical applicability of the wearable system in real-life conditions, we conducted a pilot study involving two healthy human volunteers. Sweat samples were collected during moderate physical activity and analyzed using the integrated SPR sensor embedded in the Polar Watch. Prior to measurement, the skin surface was gently cleaned with water and dried to minimize environmental contamination. The total sweat volume collected from each participant was approximately 2 mL, sufficient for triplicate testing. As shown in **Fig. 3h–i**, the reflectance spectra of sweat samples from both volunteers exhibited clear signal shifts consistent with the calibration curves established using glucose standards. Based on the relative intensity changes / primary changes (ΔI/I) and the linear relationship established in **Fig. 3g**, the glucose concentration in sweat was estimated to be 0.28 mM for volunteer A (**Fig.3h**) and 0.21mM for volunteer B (**Fig.3i**). These values fall within the commonly reported physiological range of sweat glucose (0.06–0.3 mM), further supporting the system's accuracy in biofluid sensing. Notably, the sensor response in sweat was stable over repeated measurements, and the system showed no significant drift during the 10-minute monitoring window. This indicates good reliability under dynamic, non-laboratory conditions. Additionally, no external reagents or preprocessing steps were required, highlighting the convenience and portability of the fully integrated device. These results demonstrate that the wearable SPR sensor system can perform real-time, non-invasive glucose analysis in authentic physiological environments. The successful detection of glucose in sweat with high fidelity confirms the feasibility of using this platform for continuous health monitoring and marks a significant advancement toward practical deployment in diabetes care.

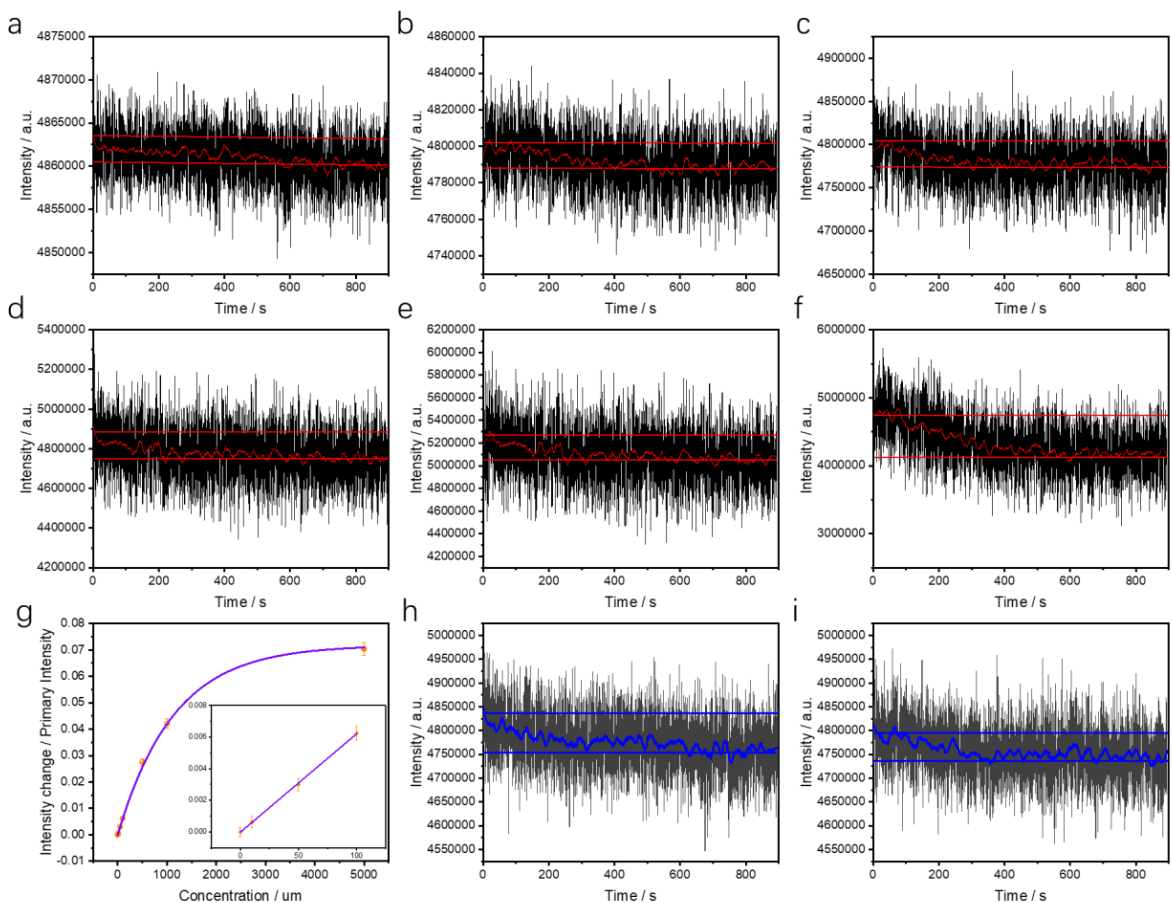

Fig. 3. (a)-(f) SPR spectra based on different glucose concentrations in functionalized silver-coated nanopillar. (g) Calibration curve of SPR reflectance intensity of functionalized silver-coated nanorods versus different glucose concentrations. (h) SPR spectroscopy of sweat from healthy volunteers A using silver nanopillar. (i) SPR spectroscopy of sweat from healthy volunteers B using silver nanopillar.

Taken together, the results presented in this study demonstrate a comprehensive strategy for overcoming key limitations in non-invasive glucose monitoring. By integrating a functionalized nanopillar substrate with optimized material composition, our system achieves high sensitivity and selectivity for glucose detection in complex sweat environments. The successful replacement of gold with silver in the SPR substrate significantly improved signal response in the red-light region, which is critical for low-power wearable applications. Moreover, the real-time wireless data transmission enabled by the wearable watch system offers a user-friendly interface for personalized health tracking. Unlike many existing wearable sensors that rely on electrochemical principles and suffer from limited selectivity or skin irritation, our optical approach provides a label-free, minimally intrusive solution with strong potential for long-term use. The dual-mode sensing capability (SERS and SPR), combined with molecularly selective functionalization via 4-MPBA, establishes a robust platform adaptable to other biomarkers beyond glucose. Future work will focus on expanding the sample size for clinical validation, improving the stability of the sensor under long-term use, and integrating advanced data analysis algorithms for intelligent health feedback. The modularity of this sensing system also opens the door to multiplexed detection and broader applications in wearable diagnostics.

## Conclusion

In this study, we developed a wearable surface plasmon resonance (SPR) sensor system based on functionalized silver-coated silicon nanowires (Ag/SiNWs) for non-invasive, real-time glucose monitoring in human sweat. By leveraging 4-mercaptophenylboronic acid (4-MPBA) as a selective molecular interface, combined with nanostructure-enhanced optical sensitivity, the platform achieved robust detection within physiologically relevant concentration ranges. The substitution of gold with silver significantly enhanced red-light SPR performance, enabling the use of a compact 638 nm laser for integration into a wearable watch. Real-world validation demonstrated the system's reliability and accuracy in sweat-based glucose detection with a detection limit of 0.052 mM. These findings highlight the feasibility of combining nanoplasmonic materials, molecular recognition chemistry, and miniaturized optics to create practical wearable biosensors. The proposed system offers a promising route toward non-invasive, user-friendly glucose monitoring and lays the groundwork for future development of multiplexed wearable diagnostic devices for personalized healthcare.

## Data availability

The data that support the findings of this study are available from the corresponding author upon reasonable request.

## Author contributions

Ling Liu: conceptualization, methodology, data curation, validation, visualization, writing-original draft and editing. Kuo Zhan: conceptualization, methodology, data curation, formal analysis, investigation, writing-review and editing. Joni.Kilpijarvi: conceptualization, methodology, data curation, project administration, resources. Matti.Kinnunen: conceptualization, methodology, data curation, project administration, resources. Artem Zhyvolozhnyi: conceptualization, methodology, data curation. Yuan zhang: conceptualization, methodology, data curation. Yang Li: writing-review and editing. Jian-an Huang: conceptualization, methodology, investigation, data curation, writing-review and editing, funding acquisition, project administration, resources.

## Acknowledgements

This research receives support from Tandem Industry Academia 2021 project (No. 312) and DigiHealth project (No. 326291), a strategic profiling project at the University of Oulu that is supported by the Academy of Finland and the University of Oulu. Ling Liu and Yuan Zhang acknowledge the China Scholarship Council for a scholarship for doctoral study at the University of Oulu. Jianan Huang thank Professor Caglar Elbuken in University of Oulu for valuable discussion about microfluidic collection of sweats.

**Reference**

1. D. R. Whiting, L. Guariguata, C. Weil and J. Shaw, *Diabetes research and clinical practice*, 2011, **94**, 311-321.
2. A. F. Amos, D. J. McCarty and P. Zimmet, *Diabetic medicine : a journal of the British Diabetic Association*, 1997, **14 Suppl 5**, S1-85.
3. H. Lee, C. Song, Y. S. Hong, M. Kim, H. R. Cho, T. Kang, K. Shin, S. H. Choi, T. Hyeon and D.-H. Kim, *Science advances*, 2017, **3**, e1601314.
4. H. Teymourian, A. Barfidokht and J. Wang, *Chemical Society reviews*, 2020, **49**, 7671-7709.
5. C. Chen, Q. Xie, D. Yang, H. Xiao, Y. Fu, Y. Tan and S. Yao, *Rsc Advances*, 2013, **3**, 4473-4491.
6. G. Cappon, M. Vettoretti, G. Sparacino and A. Facchinetti, *Diabetes & metabolism journal*, 2019, **43**, 383.
7. M. Li, L. Wang, R. Liu, J. Li, Q. Zhang, G. Shi, Y. Li, C. Hou and H. Wang, *Biosensors and Bioelectronics*, 2021, **174**, 112828.
8. Z. Sonner, E. Wilder, T. Gaillard, G. Kasting and J. Heikenfeld, *Lab on a Chip*, 2017, **17**, 2550-2560.
9. J. R. Sempionatto, I. Jeerapan, S. Krishnan and J. Wang, *Analytical chemistry*, 2019, **92**, 378-396.
10. J. Niu, S. Lin, D. Chen, Z. Wang, C. Cao, A. Gao, S. Cui, Y. Liu, Y. Hong and X. Zhi, *Small*, 2024, **20**, 2306769.
11. T. R. Ray, J. Choi, A. J. Bandodkar, S. Krishnan, P. Gutruf, L. Tian, R. Ghaffari and J. A. Rogers, *Chemical reviews*, 2019, **119**, 5461-5533.
12. S. K. Vashist, *Analytica chimica acta*, 2012, **750**, 16-27.
13. T. Saha, R. Del Caño, K. Mahato, E. De la Paz, C. Chen, S. Ding, L. Yin and J. Wang, *Chemical Reviews*, 2023, **123**, 7854-7889.
14. A. Koh, D. Kang, Y. Xue, S. Lee, R. M. Pielak, J. Kim, T. Hwang, S. Min, A. Banks and P. Bastien, *Science translational medicine*, 2016, **8**, 366ra165-366ra165.
15. D. Nakayama, Y. Takeoka, M. Watanabe and K. Kataoka, *Angewandte Chemie International Edition*, 2003, **42**, 4197-4200.
16. J. C. Pickup, F. Hussain, N. D. Evans, O. J. Rolinski and D. J. Birch, *Biosensors and Bioelectronics*, 2005, **20**, 2555-2565.
17. P. Singh, *Sensors and actuators B: Chemical*, 2016, **229**, 110-130.
18. X. Yang, Q. Wang, K. Wang, W. Tan and H. Li, *Biosensors and Bioelectronics*, 2007, **22**, 1106-1110.

19. J.-A. Huang, Y.-Q. Zhao, X.-J. Zhang, L.-F. He, T.-L. Wong, Y.-S. Chui, W.-J. Zhang and S.-T. Lee, *Nano letters*, 2013, **13**, 5039-5045.
20. H. Zafar, A. Channa, V. Jeoti and G. M. Stojanović, *Sensors*, 2022, **22**, 638.
21. R. L. McCreery, *Raman spectroscopy for chemical analysis*, John Wiley & Sons, 2005.
22. X. Sun, *Analytica Chimica Acta*, 2022, **1206**, 339226.
23. K. E. Shafer-Peltier, C. L. Haynes, M. R. Glucksberg and R. P. Van Duyne, *Journal of the American Chemical Society*, 2003, **125**, 588-593.
24. D. Yang, S. Afroosheh, J. O. Lee, H. Cho, S. Kumar, R. H. Siddique, V. Narasimhan, Y.-Z. Yoon, A. T. Zayak and H. Choo, *Analytical chemistry*, 2018, **90**, 14269-14278.
25. M. Dautta, M. Alshetaiwi, J. Escobar and P. Tseng, *Biosensors and Bioelectronics*, 2020, **151**, 112004.

# Supporting Information

**Bridging Optical Sensing and Wearable Health Monitoring: A Functionalized Plasmonic Nanopillar for Non-Invasive Sweat Glucose Detection**

Ling Liu[1], Kuo Zhan[1]*, Joni Kilpijärvi[2], Matti Kinnunen[2], Yuan Zhang[1], Mulusew Yaltaye[1], Yang Li[1], Artem Zhyvolozhnyi[3], Anatoliy Samoylenko[3], Seppo Vainio[3], Jianan Huang[1]*

[1]Research Unit of Disease Networks, Faculty of Biochemistry and Molecular Medicine University of Oulu, Oulu, Finland.

[2]Polar Electro Oy, oulu, Finland.

[3]Faculty of Biochemistry and Molecular Medicine, Disease Networks Research Unit, InfoTech Oulu, Kvantum Institute, University of Oulu, Borealis Biobank of Northern Finland, Univ. Oulu Hospital, FI-90014 Oulu, Finland.

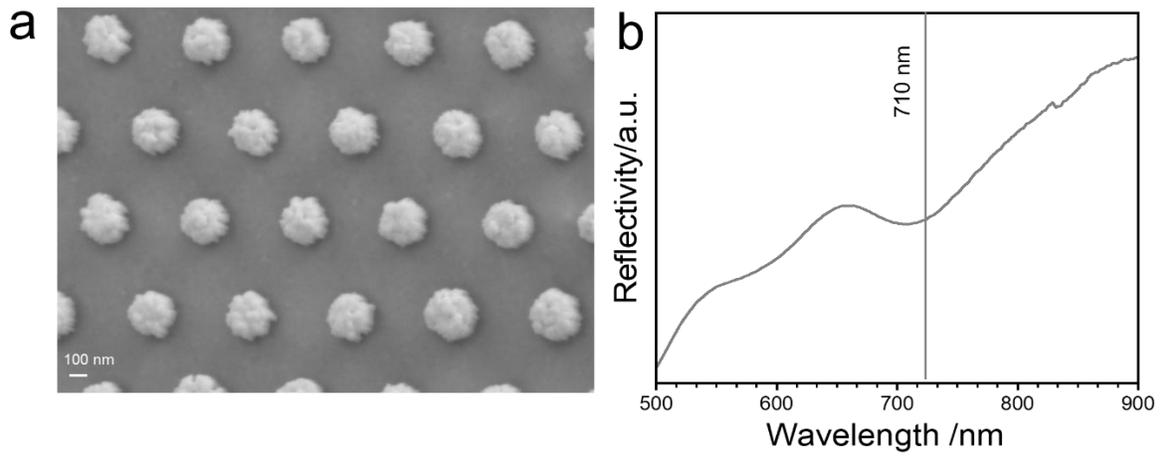

**Fig. S1** (a) SEM of gold nanopillar from a top-down angle. (b) Reflectivity spectrum of a 50 nm thick gold film coated on a 280 nm diameter and 780 nm length SiNW.

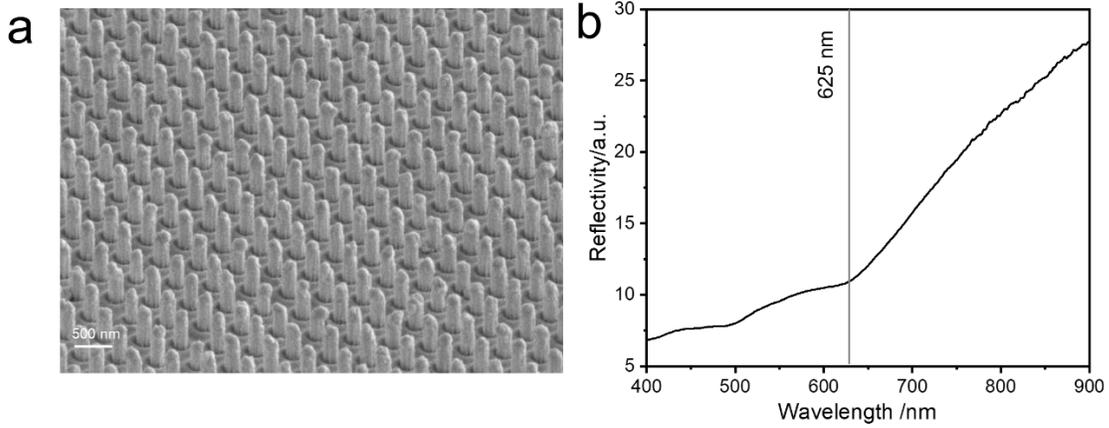

**Fig. S2** (a) Top-down SEM image of gold nanopillar tilted at an angle of 52 degrees. (b) Reflectivity spectrum of a 50 nm thick silver film coated on a 280 nm diameter and 780 nm length SiNW.

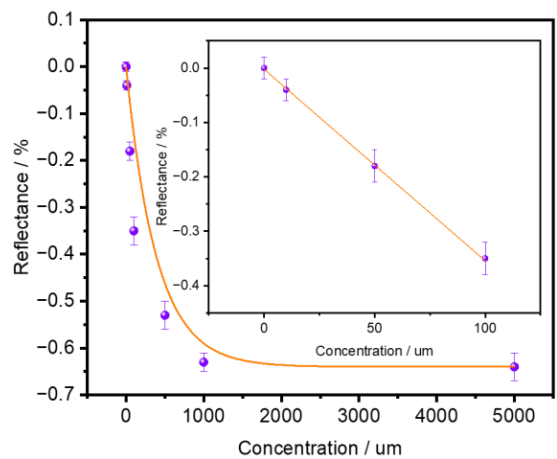

**Fig. S3** Linear relationship between SPR reflectance intensity of 4–MPBA–modified Au-coated nanopillar *versus* glucose concentration with LOD to be 0.33 mM.